\documentclass[prb,twocolumn,superscriptaddress,showpacs]{revtex4}

\usepackage{amsmath,amssymb,graphicx,bm,soul}
\usepackage[usenames,dvipsnames]{color}

\begin{document}

\title{Instability-induced formation and non-equilibrium dynamics of phase defects in polariton condensates.}

\author{T. C. H. Liew}
\affiliation{Division of Physics and Applied Physics, Nanyang Technological University 637371, Singapore}

\author{O. A. Egorov}
\affiliation{Institute of Condensed Matter Theory and Solid State Optics, Abbe
Center of Photonics, Friedrich-Schiller-Universit\"at Jena, Max-Wien-Platz
1, 07743 Jena, Germany}

\author{M. Matuszewski}
\affiliation{Instytut Fizyki Polskiej Akademii Nauk, Aleja Lotnik\'ow 32/46, PL-02-668 Warsaw, Poland}

\author{O. Kyriienko}
\affiliation{Division of Physics and Applied Physics, Nanyang Technological University 637371, Singapore}
\affiliation{QUANTOP, Danish Quantum Optics Center, Niels Bohr Institute, University of Copenhagen, Blegdamsvej 17, DK-2100 Copenhagen, Denmark}

\author{X. Ma}
\affiliation{Institute of Condensed Matter Theory and Solid State Optics, Abbe
Center of Photonics, Friedrich-Schiller-Universit\"at Jena, Max-Wien-Platz
1, 07743 Jena, Germany}

\author{E. A. Ostrovskaya}
\affiliation{Nonlinear Physics Centre, Research School of Physics and Engineering, The Australian National University, Canberra ACT 2601, Australia}

\date{January 23, 2015}

\begin{abstract}

We study, theoretically and numerically, the onset and development of modulational instability in an incoherently pumped spatially homogeneous polariton condensate. Within the framework of mean-field theory, we identify regimes of modulational instability in two cases: 1) Strong feedback between the condensate and reservoir, which may occur in scalar condensates, and 2) Parametric scattering in the presence of polarization splitting in spinor condensates. In both cases we investigate the instability induced textures in space and time including non-equilibrium dynamics of phase dislocations and vortices. In particular we discuss the mechanism of vortex destabilization and formation of spiraling waves. We also identify the presence of topological defects, which take the form of half-vortex pairs in the spinor case, giving an ``eyelet'' structure in intensity and dipole type structure in the spin polarization. In the modulationally stable parameter domains, we observe formation of the phase defects in the process of condensate formation from an initially spatially incoherent low-density state. In analogy to the Kibble-Zurek type scaling for nonequilibrium phase transitions, we find that the defect density scales with the pumping rate.

\end{abstract}

\pacs{71.36.+c,42.65.Sf,78.67.-n,03.75.Mn}


\maketitle

\section{Introduction}

The formation of complex spatio-temporal patterns and textures is a particularly intriguing phenomenon occurring in  a diverse range of physical systems.\cite{Cross1993,StaliunasBook} The patterns commonly arise in nonlinear dissipative systems driven far from equilibrium. Amongst examples of such systems, exciton-polaritons in semiconductor microcavities have emerged as a hybrid light-matter system with strongly nonlinear properties. Effects such as Bose-Einstein condensation,\cite{Kasprzak2006,Balili2007,Lai2007} superfluidity~\cite{Carusotto2004,Amo2009a,Amo2009b} and the formation of solitons~\cite{Egorov2009,Egorov2010,Egorov2011,Sich2011,Ostrovskaya2012} have been well-documented in the literature.\cite{Deng2010,Carusotto2013}

A particularly important nonlinear effect in the context of pattern formation in microcavities is the parametric scattering of resonantly excited polaritons in planar semiconductor microcavities.\cite{Savvidis2000,Tartakovskii2000,Stevenson2000,Saba2001} The pair scattering of pairs of polaritons to different states in reciprocal space allows a homogeneous polariton field to spontaneously break translational symmetry. This effect enables the formation of ordered hexagonal/triangular lattices, predicted theoretically~\cite{Saito2013,Egorov2014,Werner2014} and observed experimentally~\cite{Ardizzone2013} under different excitation conditions, as well as lattices of breathing solitons.\cite{Egorov2013}

Under non-resonant/incoherent excitation, vortex lattices were predicted to occur in harmonic traps~\cite{Keeling2008,Borgh2010} and later observed in experiments involving multiple excitation spots.\cite{Cristofolini2013} The formation of multi-lobed~\cite{Manni2011} and vortex-antivortex patterns~\cite{Manni2013} under ring shaped excitation, as well as sunflower ripples~\cite{Christmann2012} excited by a narrow pump spot have also been reported. In these examples, the translational symmetry of the system is already broken by the chosen shape of the pump-spot and/or the presence of a gradient in the potential of polaritons (as in, for example, the case of harmonic traps~\cite{Balili2007}).

In this work we consider the possibility of spontaneous breaking of translational symmetry and pattern formation in planar microcavities excited by a spatially homogeneous incoherent pump. The pumping creates a reservoir of ``hot'' exciton-like polaritons, which form a polariton condensate through a stimulated scattering process. The translational symmetry breaking is triggered by linear instability of the homogeneous condensate to spatial modulations, and the nonlinear evolution of the unstable state leads to formation of spatial patterns. We consider two different mechanisms of such \emph{modulational instability }(MI) in this system. The first one arises when the polariton condensate has a strong feedback effect on the reservoir, in the form of reservoir depletion due to stimulated scattering of reservoir excitons into polaritons. In this case, while polariton-polariton interactions are repulsive, the essentially saturable nature of exciton-polariton interactions may lead to effectively attractive nonlinearity for sufficiently low pump powers.\cite{Smirnov2014} This effective focusing nonlinearity in the system naturally leads to MI of a spatially homogeneous state, which was established in several previous studies for both quasi-1D and 2D geometry.\cite{Wouters2007,Borgh2010,Carusotto2013,Littlewood06,Malpuech14,Smirnov2014,Li14}.

Modulational instability is also known in spin-1 Bose-Einstein condensates of ultracold atoms due to parametric coupling~\cite{Robins2001,Zhang2005,Li2005,Doktorov2007} and nonlinear interactions between spin components,\cite{Matuszewski2010} as well as in 1D exciton-polariton condensates with a spin (polarization) degree of freedom.\cite{Kamchatnov2013} Although polariton systems are non-conservative and non-equilibrium, the two-component spin degree of freedom of polaritons does allow a second mechanism of MI, which works also in the defocusing regime where a strong condensate-reservoir feedback is unnecessary. A circularly polarized excitation splits the energy of the $\sigma_+$ and $\sigma_-$ states due to anisotropic interactions~\cite{Vladimirova2010,Takemura2014} occurring between the spin polarized reservoir and condensed polaritons. This splitting sets the foundation for a parametric scattering process as polaritons in an initially homogeneous state with wavevector $k=0$ on the upper spin-split branch can now scatter to degenerate non-zero wavevector states on the lower branch, reminiscent of experiments under resonant excitation in triple microcavities~\cite{Diederichs2006} or experiments in one-dimensional polariton systems.\cite{Xie2012} While such a scattering process is not strictly allowed in isotropic cavities as it would violate spin conservation, the presence of sample anisotropy, which typically causes an additional linear polarization splitting and hybridization of the $\sigma_+$ and $\sigma_-$ branches, relaxes this limitation.

By considering the stability of the steady states of the system to weak perturbations, we find the zones of MI in the two different regimes. In the scalar case, where MI is derived from  the condensate-reservoir feedback, we find that the homogeneous state breaks its  translational symmetry and forms  a turbulent state of phase dislocations, i.e., vortices. Unlike the previously studied cases, vortices do not form as the result of thermal  fluctuations~\cite{Giorgetti2007} or scattering on disorder~\cite{Liew2008}. Rather, the spatial fragmentation of the initially homogeneous condensate due to the development of MI creates multiple interference between polariton flows generated by the randomly distributed sources, which leads to the development of multiple phase dislocations, similar to the scenario previously considered for multiple pump spots~\cite{Berloff10} and highly inhomogeneous trapped polariton condensates~\cite{Borgh2010,Borgh2012}.

In the case of modulationally stable background, we show that multiple phase singularities can appear as a result of mean-field evolution of an initial white noise state, which mimics a pre-condensate state lacking spatial and phase coherence. Remarkably, formation of multiple vortices in this scenario seems to be analogous to, but not the same as, the Kibble-Zurek mechanism, which acts during the quench through a phase transition to the Bose-Einstein condensation (BEC)~\cite{Kibble1976,Zurek1985}. Indeed, the later describes the formation of boundary defects between different domains of condensate which develop an independent phase rather than inheriting it from the neighbouring spatial domains~\cite{Zurek2009,Lamporesi2013}.

We stress that the process of defect formation during nonequilibrium condensation of exciton-polaritons does not follow
the scenario of the Kibble-Zurek mechanism~\cite{Zurek1985,Dziarmaga2010,Matuszewski2014}. The main difference is that in the latter,
it is assumed that the system is initially in thermal equilibrium, and is driven out of equilibrium only in the vicinity of the phase
transition~\cite{Zurek1985,Dziarmaga2010}. The process is divided into three phases, corresponding to adiabatic-impulse-adiabatic evolution.
In nonequilibrium condensation, the system is far from equilibrium at the outset, and the transition to the quasi-equilibrium (condensed)
state occurs only after crossing the critical point.

Nevertheless, the Kibble-Zurek mechanism and the defect formation in nonequilibrium systems have much in common. In both cases, defects
are created due to symmetry breaking in separate parts of the system which cannot communicate in a finite time.
In both cases, there is a competition of two timescales existing in the system, which results in the same algebraic forms of power-law scalings
for the number of defects and their characteristic creation time~\cite{Matuszewski2014}.
In the polariton condensation case, the quench time is replaced
by the timescale of the formation of the condensate, which is controlled by the external pumping rate. We refer the reader
to Sec.~\ref{Ch:ScalingLaw} and Ref.~\onlinecite{Matuszewski2014} for the detailed description of the process.

Regardless of the mechanism of the vortex formation, either as the result of the MI development or as a result of transition to BEC, we show that the presence of the incoherent reservoir affects substantially both stability of vortices and their collective dynamics even for the case of a stable homogeneous background. As a consequence, the vortices can lose their radial symmetry and develop either into spatially localized rotating phase dislocations or into non-localized spiraling waves.

A similar situation occurs in the spinor case, although multiple branches of modulationally unstable and stable solutions are present. Defects in the spin polarization of the condensate may appear even in the modulationaly stable regime. Such structures move randomly in the microcavity plane and are composed of half-vortex~\cite{Rubo2007} half-antivortex pairs,\cite{Manni2012} exhibiting an associated dipole type spin texture. We predict that the density of vortices grows with increasing pump power similarly to the Kibble-Zurek scaling behaviour.

The paper is organised as follows. In Sec.~\ref{Ch:Model}, we describe the mathematical model of a semiconductor microcavity operating in the strong coupling regime under incoherent homogeneous optical pump of a circular polarization. Then, in Sec.~\ref{Ch:ScalarCase}, we study the stability and collective dynamics of phase dislocations in a single-component polariton condensate. Here the dynamics is mostly affected by the modulational instability originating from the strong feedback between the condensate and reservoir. In Sec.~\ref{Ch:SpinorCase}, we report a numerical analysis of the condensate dynamics in the presence of polarization splitting in spinor condensates. In Sec.~\ref{Ch:ScalingLaw}, we study the defect formation and the scaling laws for their density in analogy to the Kibble-Zurek mechanism.

\section{Theoretical Model}  \label{Ch:Model}

\noindent Let us begin by considering the incoherent excitation of a spinor polariton system, i.e., the system where the polarization degree of freedom is significant. The scalar case, which is valid when only one spin component is populated, is then easily obtained by removing one of the spin components. It is worth recalling that experiments with a circularly polarized optical pump have resulted in the excitation of a circularly polarized polariton condensate at the pump position, in both 2D~\cite{Kammann2012} and 1D~\cite{Anton2014} samples. A theoretical model can be based on the generalized Gross-Pitaevskii approach,\cite{Wouters2007} where a condensate of exciton-polaritons can be described by the wavefunctions, $\psi_+$ and $\psi_-$, of the $\sigma^+$ and $\sigma^-$ circularly polarized states, respectively:
\begin{align}
i\hbar\frac{d\psi_+}{dt}&=\left(-\frac{\hbar^2\nabla^2}{2m}+g_1n_R+\alpha_1|\psi_+|^2+\alpha_2|\psi_-|^2\right)\psi_+\notag\\
&\hspace{10mm}+i\hbar \left(r n_R-\Gamma\right)\psi_++\Delta_\mathrm{XY}\psi_-,\label{eq:GPplus}
\end{align}
\begin{align}
i\hbar\frac{d\psi_-}{dt}&=\left(-\frac{\hbar^2\nabla^2}{2m}+g_2n_R+\alpha_2|\psi_+|^2+\alpha_1|\psi_-|^2\right)\psi_-\notag\\
&\hspace{10mm}-i\hbar\Gamma\psi_-+\Delta_\mathrm{XY}\psi_+.\label{eq:GPminus}
\end{align}
Here we assume that the circularly polarized pumping creates a circularly polarized reservoir, $n_R$, with dynamics described by the rate equation:
\begin{equation}
\frac{dn_R}{dt}=-\left(\Gamma_R+r|\psi_+|^2\right)n_R+P,
\label{eq:Reservoir}
\end{equation}
where $P$ represents the pumping rate. Non-linear interactions between polaritons are characterized by $\alpha_1$ and $\alpha_2$, representing the interaction strengths between parallel and antiparallel spins~\cite{Vladimirova2010,Takemura2014}, respectively. Similar parameters, $g_1$ and $g_2$, characterize the blueshift caused by the circularly polarized reservoir. $\Gamma$ is the polariton decay rate. $\Delta_{XY}$ represents a linear polarization splitting, which has been reported in several experimental studies~\cite{Klopotowski2006,Krizhanovskii2006,Amo2009} and can take values of 0-0.2 meV.\cite{Klopotowski2006,Comitti2014} Even larger values can be expected by the application of magnetic fields (in Voight configuration).\\

We note that in the limit $\Gamma_R\gg\Gamma$, it is possible to proceed by adiabatic elimination of the reservoir dynamics.\cite{Borgh2010,Larre2013} However we do not make such an approximation here since the reservoir dynamics is important for our further analysis.

\section{Non-equilibrium Dynamics in the Scalar Case} \label{Ch:ScalarCase}
\subsection{Modulational instability of the homogeneous steady state.}
\begin{figure}
\includegraphics[width=0.95\linewidth]{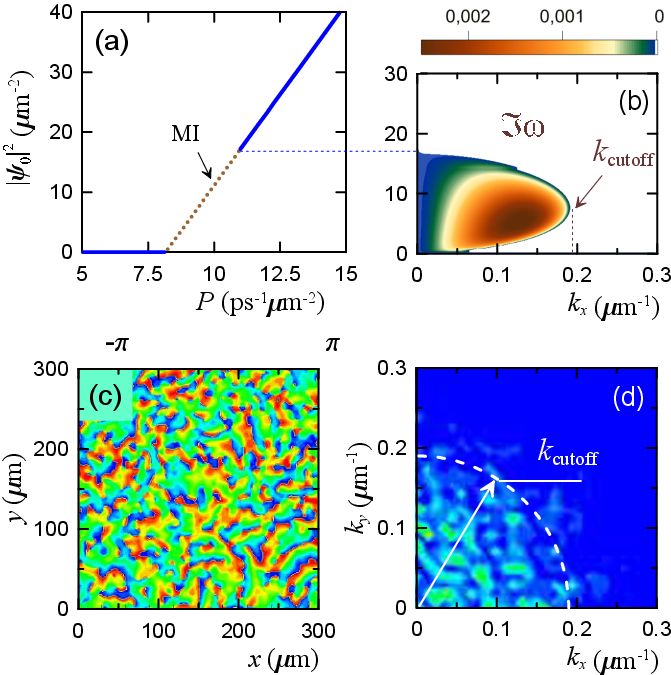}
\caption{ (Color online) Steady-state solutions of the model~(\ref{eq:GPplus}-\ref{eq:Reservoir}) and their stability for a circularly polarized pump ($\Delta_\mathrm{XY}=0$).
(a) The homogeneous solution (HS) ($|\psi_0|^2$) as functions of pump $P$. The dotted line depicts the modulational instability (MI). (b) Growth rates [$\Im m  \omega$ (ps$^{-1}$)] of the small perturbations (around $|\psi_0|^2$) as a function of momentum $k$ and HS density $|\psi_0|^2$. The vertical axis is the same as in (a) and the horizontal dashed (blue) line connecting (a) and (b) indicates that the boundary of modulational instability occurs for the same maximum homogeneous density. Snapshots of the condensate phase in real space (c) and the density in two-dimensional reciprocal space (d) within the MI domain of a HS for pump $P=9$~ps$^{-1} \mu m^{-2} $. $k_\mathrm{cutoff}$ can be identified as the magnitude of the in-plane wavevector at which all unstable modes disappear. Parameters: $\alpha_1=6\times10^{-3}$~meV$\mu$m$^2$, $g_1=4\alpha_1$, $\Gamma=0.165$~ps$^{-1}$, $r=0.01$~ps$^{-1}\mu$m$^2$, $\Gamma_R=0.5$~ps$^{-1}$. The polariton effective mass was taken as $\times10^{-4}$ of the free electron mass.}
\label{fig:SteadyState_Scalar}
\end{figure}
If we consider a circularly polarized pump in a microcavity with negligible polarization splitting ($\Delta_\mathrm{XY}=0$) then the model can be reduced to a scalar one ($\psi=\psi_+$) without the second polarization component ($\psi_-=0$). First, we study a scalar steady-state homogeneous solution (HS) of the system of Eqs. (\ref{eq:GPplus})-(\ref{eq:Reservoir}) and discuss its stability. For the sake of generality we allow also the HSs with nonzero transversal momenta $k_0\neq0$, which have the form of travelling waves
\begin{equation}
{\psi _{ \rm   hs}}(x,t) = {\psi _0}{e^{ - i\,{\mu}({k_0},\left| {{\psi _0}} \right|)t + i{k_0}x}},
 \label{eq:HomSol}
\end{equation}
where the condensate energy is given by $\hbar \mu^{}\left( {{k_0},{{\left| {{\psi _0}} \right|}}} \right) = \left( {{{{\hbar ^2}} \mathord{\left/
 {\vphantom {{{\hbar ^2}} {2m}}} \right.
 \kern-\nulldelimiterspace} {2m}}} \right)k_0^2 + {\alpha _1}{\left| {{\psi _0}} \right|^2} + {g_1}{n_{R0}}$.
The HS becomes nontrivial provided that the external pump compensates for all losses and overcomes the threshold value:\cite{Wouters2007} ${P_{th}} = \Gamma \Gamma _R/r$. The coherent exciton-polariton density and incoherent reservoir density are given by $\left| \psi _0 \right|^2 =  (P - P_{th})/\Gamma$ and $n_{R0} = \Gamma  /r$, respectively.

The linear stability analysis of the homogeneous steady state of our scalar system and its modifications has been previously performed by many authors.\cite{Wouters2007,Carusotto2013,Littlewood06,Malpuech14,Smirnov2014,Li14} For our choice of the system parameters, the linear stability analysis shows that the HS becomes modulationally (dynamically) unstable within a pump interval just above the threshold value of the pump $P_{th}$ [Fig.~\ref{fig:SteadyState_Scalar}(a)] (details of the analysis are given in Appendix~\ref{Ch:AppendixStability}).
This MI is associated with the parametrical generation of field components with nonzero momenta $k$. Figure ~\ref{fig:SteadyState_Scalar}(b) presents the linear growth rate $\Im m \omega(k)>0$ of the unstable perturbations as a function of their momenta $k$.

It has been shown recently~\cite{Smirnov2014} that this MI is associated with the effective attractive nonlinearity induced by the saturation of the incoherent reservoir. Indeed, based on the intuition gained from paradigm nonlinear models, such as the Schr\"{o}dinger equation with a Kerr-nonlinearity, one expects that the existence of the MI requires a focusing nonlinearity.\cite{Saito2001,AlKhawaja2002,Salasnich2003,Carr2004} However, owing to repulsive interactions between excitons, the nonlinear behaviour of an exciton-polariton condensate is akin to that of optical waves in a defocusing media. This seeming contradiction clearly elucidates the influence of the open-dissipative nature of the system on the nonlinear behaviour,\cite{Smirnov2014} which requires inclusion of an incoherent reservoir of ``hot'' excitons. To illustrate this influence we consider a nonlinear energy shift induced by both the coherent exciton-polaritons and the incoherent reservoir
\begin{equation}
{\hbar\mu_{nl}}\left( {{{\left| \psi  \right|}^2}} \right) = {\alpha _1}{\left| \psi  \right|^2} + {{g}_1}\frac{{{P}}}{{{\Gamma _R} + r{{\left| \psi  \right|}^2}}}. \label{eq:NonlShift}
\end{equation}
We note that the polariton density $|\psi|^2$ corresponds to a steady-state solution which, in general, is not necessarily given by the homogeneous value $|\psi_0|^2$. The reservoir intensity $n_R$ follows this steady-state solution $|\psi|^2$.
\begin{figure}
\includegraphics[width=0.9\linewidth]{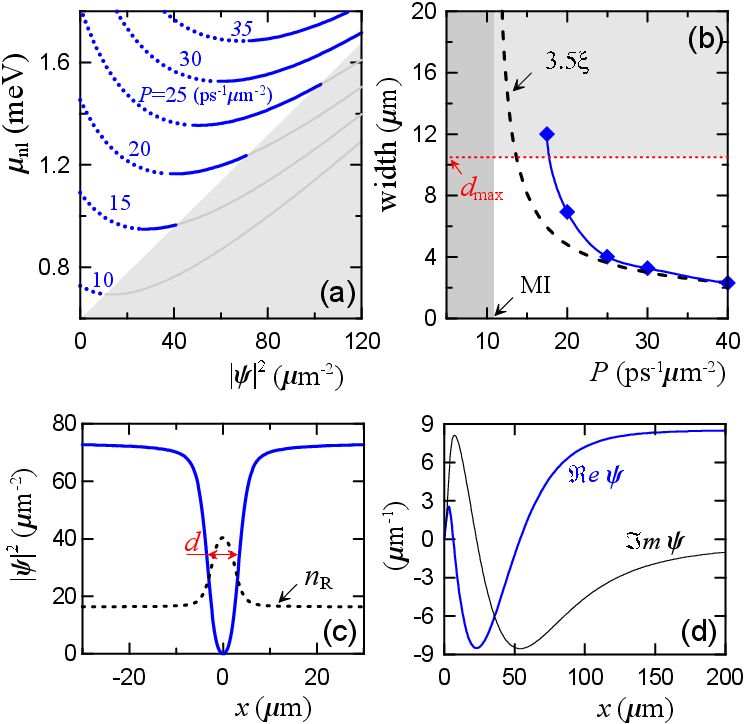}
\caption{ (Color online) (a) Effective nonlinear energy shift $\hbar\mu_{nl}$ versus the condensate  density, $|\psi|^2$, for different pump $P$. Dashed lines (with  the negative slope) represent an effectively focusing nonlinear response (or nonlinear red shift). The grey triangle represents the intensities $|\psi|^2$ exceeding values given by the steady-state homogeneous solutions ($|\psi_0|^2$). (b) Vortex width (FWHM) vs. the pump $P$. The dashed line represents the width approximated by the value $\approx3.5\xi$. $d_{max}$ depicts the maximal diameter allowed for stable vortices, see Eq.~(\ref{eq:DiametrMax}). (c) The radial dependence of the condensate and reservoir density within the vortex profile for $P=20$~ps$^{-1} \mu m^{-2}$. (d) The real and imaginary parts of the $\psi$ within the vortex profile given in (c). Other parameters are the same as in Fig.~\ref{fig:SteadyState_Scalar}. }
\label{fig:NonlinShift}
\end{figure}
The first term of Eq.~(\ref{eq:NonlShift}) describes the blue shift originating from the repulsive exciton-exciton interaction, whereas the second term describes the saturation of the reservoir. We define the effective nonlinearity coefficient as $g_{ \rm  eff} = \hbar{{\partial {\mu_{nl}}} \mathord{\left/ {\vphantom {{\partial {\mu _{nl}}} \partial }} \right. \kern-\nulldelimiterspace} \partial }{\left| \psi  \right|^2}$. Then the nonlinear response is effectively focusing provided that the coefficient is negative ${g_{ \rm  eff}}<0$. Otherwise, if ${g_{ \rm  eff}}>0$ the nonlinear response is defocusing or repulsive [see thick solid lines in the Fig.~\ref{fig:NonlinShift}(a)].
In the vicinity of the steady-state HS with intensity $\left| \psi_0  \right|^2$ this coefficient takes the form  (cf. Eq.~(36) in Ref.[\onlinecite{Smirnov2014}]):
\begin{equation}
 {g_{ \rm  eff}} = {\alpha _1} \left(1- \frac{{{g_1}{\Gamma ^2}}}{{\alpha _1rP}} \right). \label{eq:NonlCoef}
\end{equation}
The effective nonlinear coefficient [Eq.~(\ref{eq:NonlCoef})] changes sign for a pump value $ {P_{MI}} = {{{g_1}{\Gamma ^2}}}/{{\alpha _1 r}}$. 
It means that in the vicinity of the HS the nonlinear response changes character from effectively focusing to effectively defocusing. As a result the modulationally unstable HS becomes stable exactly at this point $P=P_{MI}$. The condition $P_{MI}>P_{th}$ gives a general criterion for the appearance of MI derived in Ref.~[\onlinecite{Smirnov2014}]:
\begin{equation}
\frac{{{g_1}\Gamma }}{{{\alpha _1}{\Gamma _R}}} > 1.    \label{eq:MI_Crit}
\end{equation}

Within the MI interval ($P_{th}<P<P_{MI}$) indicated in Fig.~\ref{fig:SteadyState_Scalar}(a), the HS experiences spontaneous translational symmetry breaking, resulting in the formation of non-uniform turbulent states of the condensate.

The result of a direct numerical calculation is shown in Fig.~\ref{fig:SteadyState_Scalar}(c). In these calculations and those presented throughout the manuscript we use a square grid with at least $128\times128$ points covering the plot area. We make use of an adaptive step Adams-Bashforth-Moulton procedure, which was previously found to be consistent with fixed step methods with an integration step of $0.01$ps. To double check the results of our numerical simulations we repeated them with higher precision using up to $1024\times1024$ grid points. Unless stated otherwise, periodic boundary conditions are applied.

The strongly non-equilibrium state in Fig.~\ref{fig:SteadyState_Scalar}(c) includes one- and two-dimensional phase dislocations which move chaotically and overlap. Therefore this dynamics can be characterized as a ``strong'' turbulence regime with overlapping defects (see Ref.~[\onlinecite{Berloff10}] and references therein). Two-dimensional Fourier transformation shows that the most part of the spatial momenta are bounded within the ring with the radius given by the cutoff momenta of the unstable modes $k_{ \rm  cutoff}$ [Fig.~\ref{fig:SteadyState_Scalar}(d)]. Note that direct numerical simulations of the model~(\ref{eq:GPplus},\ref{eq:Reservoir}) for different initial conditions did not reveal the formation of stationary periodical patterns, known for the coherently pumped polaritonic systems~\cite{Saito2013, Ardizzone2013,Egorov2014,Werner2014}. 
This is due to destabilization caused by fluctuations in the exciton reservoir. The modulational instability discussed later in the spinor case (section IV) does not depend on having a dynamic reservoir and can largely be reproduced assuming a static reservoir~\cite{Keeling2008}. This allows an explicit testing of the effect of the dynamic reservoir in the spinor case, where we find that periodic patterns are possible with a static reservoir but are prevented as soon as the reservoir density is allowed to evolve spatiotemporally. We assume that it is the freedom for density fluctuations to appear in the reservoir that lead to the disruption of regular patterns in the condensate also in the scalar case.


\subsection{Single vortices in the dynamically stable regime.} \label{sec:scalarfocusing}

The reservoir contribution becomes negligible in the limit of very strong pump $P \gg P_{th}$. We expect that in this case the nonlinear dynamics is very similar to that known for the conservative systems, including the formation of the stable phase dislocations and vortices. Indeed the scalar version of the equations possesses vortex solutions within the stability interval of the HS for $P>P_{MI}$ [see Figs.~\ref{fig:NonlinShift}(b) and (c)]. Similar to the conservative case, one can define a characteristic length or an effective healing length in the vicinity of the HS $\left| \psi_0 \right|^2$:
\begin{equation}
 \xi(P)  = \frac{\hbar }{{\sqrt {2m{\kern 1pt} {\kern 1pt} {g_{ \rm   eff}}{{\left| {{\psi _0}(P)} \right|}^2}} }}. \label{eq:Healing}
\end{equation}
The healing length is a typical length scale over which $\psi$ can change significantly. It also gives the typical size of the vortices. More precisely the \emph{full width at half maximum }(FWHM) of the vortices is approximately given by $d\thickapprox3.5\xi$ at least for a strong enough pumping [$P\gg P_{th}$ in Fig.~\ref{fig:NonlinShift}(b)]. It is worth to mention that the spatial oscillations of the real and imaginary parts of the amplitude profiles of the vortices go substantially beyond the healing length [Fig.~\ref{fig:NonlinShift}(d)], especially for a moderate pumping in the vicinity of the MI. This indicates permanent energy and polariton exchange within the vortex profile. The presence of these intrinsic fluxes essentially changes the interaction dynamics between vortices \cite{FraserNJP2009} providing a purely dissipative mechanism for their mutual repulsion.

The radial symmetry of the vortex phase is broken also for an inhomogeneous pump, for instance, a Gaussian pump~\cite{Borgh2010,Ostrovskaya2012}. As a result, the vortices have spiraling phases indicating again the permanent exchange of particles between different points within the resonator plane.


The stability analysis, discussed in the context of the HSs, can also be applied to the vortices, at least in the limit of very broad states. Indeed, following the healing length $\xi(P)$, the vortex width increases for smaller values of the pump and diverges in the vicinity of the MI [$P=P_{MI}$ in Fig.~\ref{fig:NonlinShift}(b)]. The formal condition for the effective focusing nonlinearity ($g_{  \rm    eff}({{\left| \psi  \right|}^2})<0$) is always satisfied within the vortex profile close enough to its core. Therefore the steady-state condensate on a circle with a fixed radius $\varrho_c \equiv \sqrt{x^{2}+y^{2}}$ and density ${{\left| \psi(\varrho_c)  \right|}^2}$ can become unstable against spatially modulated perturbations. The periodical boundary conditions on the circle restrict the number of available momenta to the values $k_c\simeq n/\varrho_c$, where $n$ is an integer. In the limit of large radius $\varrho_c\rightarrow\infty$ the values of the unstable momenta can be approximated by those values calculated for the HSs [see Fig.~\ref{fig:SteadyState_Scalar}(b)]. Therefore the polariton condensate becomes unstable provided that at least one of the available momenta ($k_c\simeq n/\varrho_c$) is smaller than the cut-off value $k_{ \rm  cutoff}\approx0.19 \mu m^{-1} $ for MI. This allows estimation of the vortex diameter $d_{max}$ where the instability just sets in, i.e., $k_{ \rm  cutoff}=2/d_{max}$. Therefore the vortices are stable provided that their diameters do not exceed the maximal value given by
\begin{equation}
d_{max}\simeq 2 k_{ \rm  cutoff}^{-1}. \label{eq:DiametrMax}
\end{equation}
It is shown in Fig.~\ref{fig:NonlinShift}(b) that the vortex diameter exceeds the maximal value $d_{max}$ within the pump interval $P_{MI}<P \lesssim 18$~ps$^{-1} \mu m^{-2}$ in the vicinity of the MI domain. As we will see below (in the next subsection~\ref{sec:scalarnonlinear}) they become unstable and lose their radial symmetry.

\subsection{Collective dynamics of vortices}\label{sec:scalarnonlinear}

Beyond the MI instability interval for $P>P_{MI}$ the non-trivial HS solution is stable. 
However, at the onset of condensation when it passes into the mean-field regime, the polariton state is spatially incoherent, with random phase. This leads to the generation of phase defects as the system goes through the condensate transition, in analogy to the Kibble-Zurek theory. To go beyond the mean-field approximation and describe first and second order spatial coherences when crossing the condensation threshold, one can make use of stochastic classical field approaches.\cite{Wouters2008,Wouters2009} These describe the pre-condensate state as an ensemble of fluctuating white noise states (governed by a stochastic Gross-Pitaevskii equation). The projection onto the classical condensate state upon condensation selects a particular realization of the noise. After the condensate has formed we assume that fluctuations are weak in comparison to the condensate mean-field and can be neglected. This approach reproduces the typical establishment and coherent evolution of topological defects in polariton condensate experiments~\cite{Lagoudakis2011,Manni2012}. We note that the account of fluctuations throughout the evolution would be important for describing accurately spectral polariton properties or the polariton photoluminescence below threshold~\cite{Wouters2008}. Fluctuations can in principle shift the phase boundaries between stable and unstable regions~\cite{Johne2009}, however, these shifts are expected to be limited and have little further effect on the dynamics.
\begin{figure}
\includegraphics[width=1.0\linewidth]{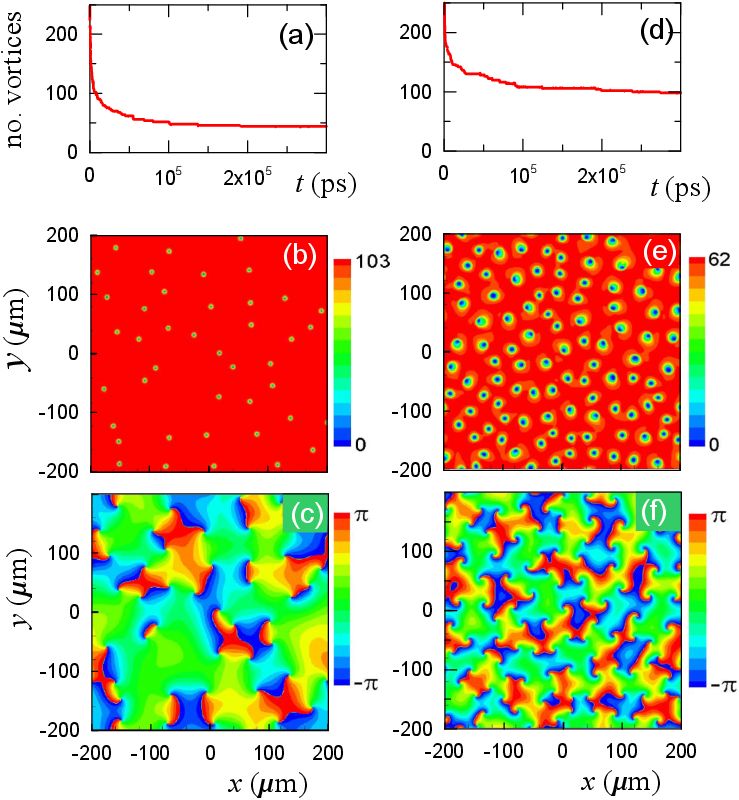}
\caption{Nonlinear dynamics of the condensate beyond the MI threshold for (a,b,c) $P=25$~ps$^{-1} \mu m^{-2}$ and (d,e,f) $P=17.5$~ps$^{-1} \mu m^{-2}$. (a,d) Long-time evolution dynamics of the dislocations number. (b,e) Snapshots of the condensate density ($\mu m^{-2}$) profiles. (c,f) Snapshots of the condensate phase profiles. See also the Supplemental Materials \cite{SM_UnstableVortices}. Parameters are the same as in Fig.~\ref{fig:SteadyState_Scalar}. }
\label{fig:VortexDynamics_Scalar}
\end{figure}

For a pumping substantially above the MI threshold the dynamics is dominated by the defocusing nonlinearity of a pure condensate ($\alpha$). The influence of dissipative dynamics of the reservoir becomes less important. Using the scalar version of the governing equations~(\ref{eq:GPplus}), (\ref{eq:Reservoir}) we calculated the condensate dynamics starting with a spatially incoherent state given by a small-amplitude white noise. Our initial condition mimics a particular realisation of the stochastic pre-condensate state. Similar to the conservative limit we observed the formation of a spatially coherent condensate accompanied by the spontaneous formation of the vortices which move chaotically and interact with each other. Two vortices with equal (opposite) topological charges repel (attract) each other. Two attracting vortices mutually annihilate if the distance between them becomes smaller than the healing length. Therefore, as discussed in Ref.~[\onlinecite{Berloff10}], the whole number of vortices drops gradually with time and approaches zero, at least for a very strong pump $P>40$~ps$^{-1} \mu m^{-2} $.

In contrast, for a weaker pump, the number of phase singularities converges eventually to some constant value indicating the formation of a coherent state with a finite number of dislocations, i.e., a superfluid turbulence [Fig.~\ref{fig:VortexDynamics_Scalar}(a)]. A snapshot of the intensity and phase profiles shows a state of well distinguishable vortices [Figs.~\ref{fig:VortexDynamics_Scalar}(b),(c)]. The vortices move chaotically, interacting with their neighbours, and, in general, sustain a dynamical equilibrium. It is remarkable that the average separations between nearest vortices remains more or less constant for this particular interval of pump values. This means that there exists some equilibrium distance between vortices. Apparently the influence of the dissipative effects and the condensate flows (mentioned in the previous subsection~\ref{sec:scalarfocusing}) are substantial.\cite{FraserNJP2009} The out-going condensate flows from the vortex centers hinder attraction between vortices and their annihilation. For even weaker pump the dissipative effects become stronger and, as a consequence, the average distance between vortices under dynamical equilibrium becomes even smaller [Figs.~\ref{fig:VortexDynamics_Scalar}(d),(e) and (f)]. We note that this dynamical equilibrium forms over a long time scale exceeding hundreds of nanoseconds. Therefore the Kibble-Zurek-like scaling law does not describe the number of vortices in this regime (see Sec.~\ref{Ch:ScalingLaw} below).

Even though the HSs are stable, the nonlinear dynamics of vortices is strongly affected by the reservoir saturation dynamics. The numerical modelling shows that the vortices themselves become unstable and develop into radially asymmetric rotating structures, as can be seen on the snapshot profiles in Figs.~\ref{fig:VortexDynamics_Scalar}(e),(f) (a movie showing the time dynamics is available in the Supplemental Material \cite{SM_UnstableVortices}). This is in agreement with the destabilization scenario for the vortices discussed in the previous subsection. Indeed the vortex size exceeds the maximal diameter $d_{max}$ given by Eq.~(\ref{eq:DiametrMax}) and therefore becomes unstable.

\subsection{Formation of spiraling waves} \label{sec:scalarvortices}
\begin{figure}
\includegraphics[width=1.0\linewidth]{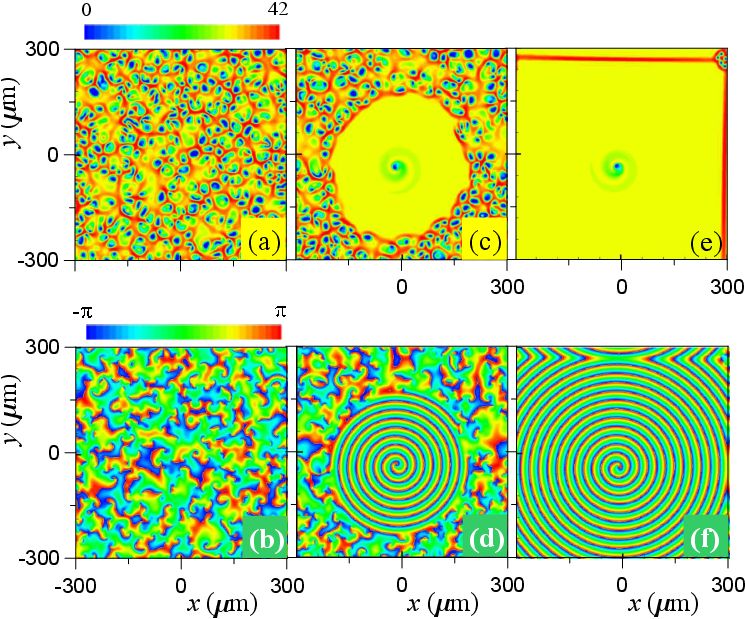}
\caption{(Color online) Snapshots of the condensate intensities (a,c,e) and phase profiles (b,d,f) for different time points and $P=13.5$~ps$^{-1} \mu m^{-2}$. (a,b) Initial turbulent state of the condensate at $t=1000$~ps. (c,d) Onset of a spiraling wave at $t=5000$~ps. (e,f) The spiraling waves at $t=11000$ps. The rotation period of the spiraling wave is $144$ps.  See also the Supplemental Materials \cite{SM_UnstableVortices}. Other parameters are the same as in Fig.~\ref{fig:SteadyState_Scalar}.}
\label{fig:Dynamics_SpiralWaves}
\end{figure}

In the vicinity of the MI the vortices are strongly unstable and the initial noise first develops into non-uniform dynamical states similar to those which appear for the modulationally unstable background [Figs.~\ref{fig:Dynamics_SpiralWaves}(a,b)]. In this ``strong'' turbulence regime~\cite{Berloff10} the characteristic distance between vortices is substantially smaller than their typical core size, so the vortices are not structured and the chaotic behavior is seen on the level of a single vortex. However, after some sufficiently long time of about several thousand polariton lifetimes, the system switches spontaneously into a more regular regime, characterized by the formation of a single spiraling topological dislocation [Fig.~\ref{fig:Dynamics_SpiralWaves}] (a movie showing the time dynamics is available in the Supplemental Material \cite{SM_UnstableVortices}). This spiraling topological  state drives away other phase dislocations in the system and covers eventually the whole computational window provided that $P_{MI}<P\lesssim 16$~ps$^{-1} \mu m^{-2}$. Similar spiraling waves are known for other non-equilibrium dissipative systems.\cite{Aranson1992,Huber1992} In the simplest case they are solutions of the complex Ginzburg-Landau equation.

It is worth to articulate the differences between the spiraling topological states and the vortices. First, there is no rotational symmetry in the profiles [Figs.~\ref{fig:Dynamics_SpiralWaves}]. Second, these topological states are not stationary and experience uniform rotation of the density with a typical rotation period of about $144$ps. Third, far from the center the profile converges to the homogeneous traveling wave solution given by Eq.~(\ref{eq:HomSol}) with  the amplitude $\psi_0$ and a nontrivial momentum $k_0$. Apparently this solution is characterized by a permanent radial flux of exciton-polaritons from the vortex center to the periphery and therefore resembles a point-like source of ring waves. We note that these spiraling waves can exist only in nonequilibrium dissipative systems.

An important question remains whether the spiraling state with a non-zero orbital angular momentum can emerge from the initially non-rotating turbulent state. Indeed, according to the conservation law of the total orbital angular momentum, the phase dislocations appear in pairs which is also valid for the turbulent state considered here. Apparently the local intensity fluctuations break the symmetry between the two dislocations within a pair in such a way that only one of them develops into the spiraling wave. The periodical boundary conditions (in $x$ and $y$ directions) were used in our numerical modelling. However, we have confirmed that the spiraling waves appear with the same probability independently on the computational window size and particular realisations of the initial seeding noise.
Moreover, since some phase dislocations are always present [see Figs.~\ref{fig:Dynamics_SpiralWaves}(e) and (f)], the total orbital angular momentum of the condensate within the computational window remains zero. We performed additional numerical simulations of the condensate dynamics under a localized pump with a ``flat-top'' shape in the form of a super-Gaussian intensity distribution. It turned out that the spiraling waves appear also for the localized pump where the condensate density vanishes at the boundaries of the computational window. These calculations serve as a solid proof of the existence of the spiraling waves independently of a particular choice of the numerical boundary conditions.

In general the out-going radiation from the center of the topological solution repels the local inhomogeneities of the profile and other topological solutions.
This gives an additional purely dissipative mechanism which enforces a long range spatial coherence in non-equilibrium systems operating in the regime of ``strong'' turbulence.

\section{Non-equilibrium dynamics in the Spinor Case} \label{Ch:SpinorCase}

\subsection{Stability of Homogeneous States}

In the presence of non-zero polarization splitting ($\Delta_\mathrm{XY}\neq0$), the population of $\sigma^-$ component appears even for a fully polarized $\sigma^+$ pumping. To emphasize the difference to the nonlinear dynamics discussed above (see Sec.~\ref{Ch:ScalarCase}) we consider system parameters which do not satisfy the MI criteria~(\ref{eq:MI_Crit}) and, therefore, guarantee the stability of homogeneous solutions in the scalar limit.

Homogeneous stationary solutions can be found by substituting trial solutions in the form $\psi_\pm(t)=\psi_\pm e^{-i\mu t}$ into Eqs.~(\ref{eq:GPplus}) and (\ref{eq:GPminus}). This gives four stationary equations for real and imaginary parts of the amplitudes. These are supplemented by the requirement that the time derivative in Eq.~(\ref{eq:Reservoir}) vanishes for a steady state, that is, $|\psi_+|^2=\frac{1}{r}\left(\frac{P}{n_R}-\Gamma_R\right)$. Noting that the phase reference of the system can be freely chosen, we can for simplicity set $\Im m\left\{\psi_-\right\}=0$, which allows one to find the relation $\Re e\left\{\psi_-\right\}=\frac{\Delta_{XY}}{\Gamma}\Im m\left\{\psi_+\right\}$ from one of the four stationary equations. The remaining three equations can be solved for the remaining unknown quantities: $\Re e\left\{\psi_+\right\}$, $\Im m\left\{\psi_+\right\}$ and $\mu$.

\begin{figure}
\includegraphics[width=1.0\linewidth]{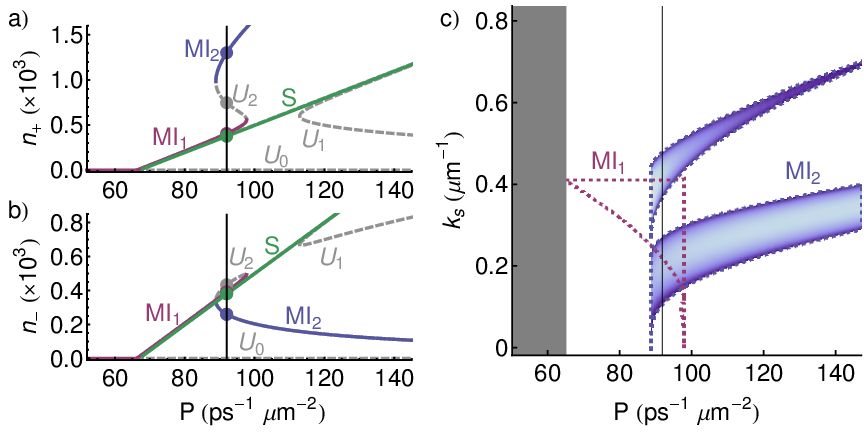}
\caption{(Color online) Stationary solutions to Eqs.~\ref{eq:GPplus}-\ref{eq:Reservoir}. (a) and (b) show the dependence of the $\sigma_+$ and $\sigma_-$ polarized polariton populations on the pumping strength, respectively. The different branches are labelled stable (S), MI, and single-mode unstable (U). (c) Parametrically unstable zones of the $\mathrm{MI}_1$ and $\mathrm{MI}_2$ branches. The shading illustrates the gain of the instability (given by the imaginary part of $\omega$) for the $\mathrm{MI}_2$ solution. The grey rectangle denotes a region below condensation threshold where $\psi_\pm=0$. The vertical line represents the pump intensity considered in Figs.~\ref{fig:spinpattern} and Fig.~\ref{fig:spinorspectrum}. Parameters: $\alpha_1=1.55\times10^{-4}$~meV$\mu$m$^2$, $\alpha_2=-0.1\alpha_1$, $g_1=\alpha_1$, $g_2=\alpha_2$, $\Gamma=0.033$~ps$^{-1}$, $r=0.01$~ps$^{-1}\mu$m$^2$, $\Gamma_R=10$~ps$^{-1}$, $\Delta_{XY}=0.1$~meV, $P=92$~ps$^{-1}\mu$m$^{-2}$. The polariton effective mass was taken as $5\times10^{-5}$ of the free electron mass.}
\label{fig:spinor}
\end{figure}

The dependence of the stationary HSs on the pumping power is shown in Figs.~\ref{fig:spinor}(a) and (b). Linear stability of the steady states can be determined by the standard extension of the Bogoliubov-de Gennes analysis \cite{Wouters2007,Carusotto2013,Littlewood06,Malpuech14,Smirnov2014,Li14,Kamchatnov2013} by considering perturbations in the polariton and reservoir fields of the form $\delta\psi_\pm=u_\pm e^{i(kx-\omega t)}+v^*_\pm e^{-i(kx-i\omega^* t)}$ and $\delta n_R=w\left(e^{i(kx-\omega t)}+e^{-i(kx-\omega^* t)}\right)$, respectively. The details of the derivation can be found in Appendix~\ref{Ch:AppendixStability}.

The stationary states labelled $U_1$ and $U_2$ in Fig.~\ref{fig:spinor}(a,b) are unstable to fluctuations with $k=0$. This ``single-mode instability'' indicates that even in a confined system (e.g., micropillar) the homogeneous steady state would be linearly unstable in the Lyapunov sense and any spatially homogeneous perturbation would grow. The stationary states labelled ${\rm MI}_1$ and ${\rm MI}_2$ are unstable to spatial modulations with non-zero wavevectors $k$. These states would be stable in a confined system, with no spatial degrees of freedom, however, in a 2D planar system the states ${\rm MI}_1$ and ${\rm MI}_2$ undergo parametric scattering. The state labelled $S$ is fully stable, as the imaginary part of $E=\hbar\omega$ remains negative for all wavevectors (see Appendix ~\ref{Ch:AppendixStability}).

The ``S''-shape of the curves ${\rm MI}_1$, ${\rm MI}_2$ and $U_2$ is characteristic of multistability, which is a common feature of resonantly excited microcavities~\cite{Gippius2004,Baas2004,Whittaker2005,Bajoni2008} but less studied under the non-resonant or incoherent excitation~\cite{Kyriienko2014} that we consider here. While multistability is strictly only present in the confined system, since the ${\rm MI}_1$ and ${\rm MI}_2$ states are unstable in the presence of spatial degrees of freedom, they can still give rise to different (non-stationary) configurations under the same excitation conditions.

Fig.~\ref{fig:spinor}(c) shows the regions of MI in the system when the $\mathrm{MI}_1$ or $\mathrm{MI}_2$ branches are excited. The $\mathrm{MI}_1$ branch begins once the threshold for polariton condensation is passed (indicated by the grey rectangle), such that only weak pump intensities are needed to see the effects of MI.

\subsection{Spin Textures due to Modulational Instability}

\begin{figure}[tbh]
\includegraphics[width=1.0\linewidth]{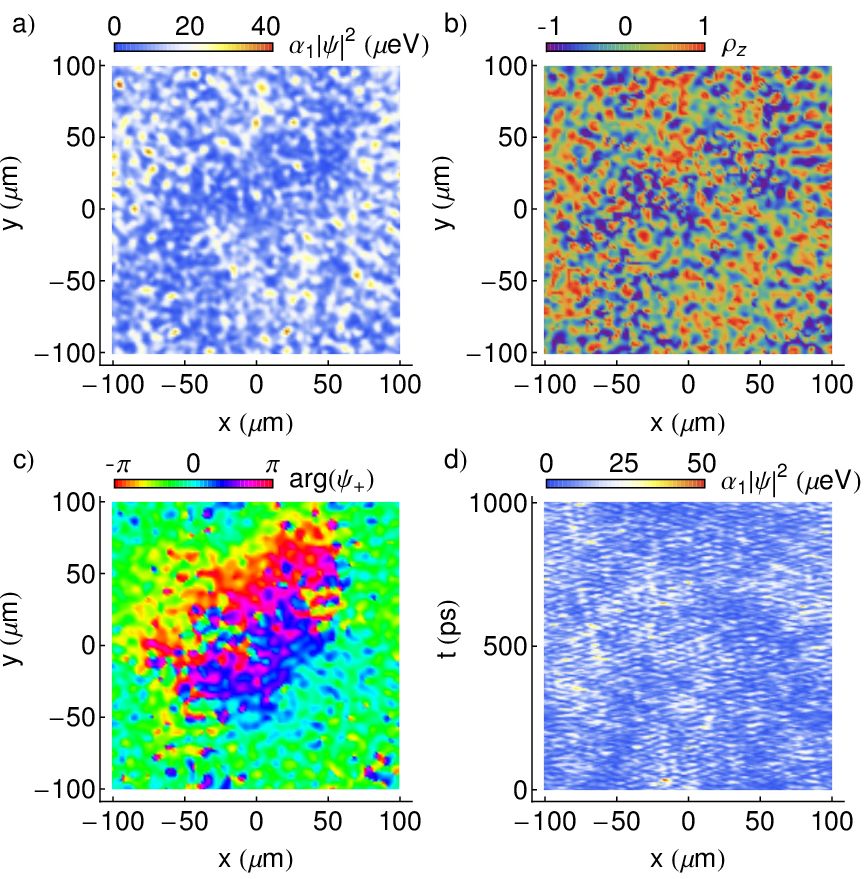}
\caption{(Color online) Spinor MI regime. (a) Density pattern of a polariton condensate under homogeneous incoherent pumping. (b) Distribution of the circular polarization degree, $\rho_z=\left(|\psi_+|^2-|\psi_-|^2\right)/\left(|\psi_+|^2+|\psi_-|^2\right)$. (c) phase of the $\psi_+$ polariton field (the $\psi_-$ component (not shown) has a similar dependence). (d) Time evolution of the polariton intensity along a slice in real space. Parameters were the same as in Fig.~\ref{fig:spinor} (a small pump power was chosen so as to excite only the $\mathrm{MI}_1$ branch).}
\label{fig:spinMI}
\end{figure}

Due to the presence of MI, we can expect the fragmentation of the homogeneous density of the condensate and spontaneous formation of spin textures, even in the presence of homogeneous pumping. Solving Eqs.~(\ref{eq:GPplus})-(\ref{eq:Reservoir}) numerically when the system is excited just above threshold on the $\mathrm{MI}_1$ branch, we obtain the spin texture shown in Fig.~\ref{fig:spinMI}. In analogy to the scalar case, the texture comprises a simultaneous modulation of the intensity and phase in the system, which oscillate with multiple frequencies. In addition to the scalar case, there is also an appearance of a non-uniform polarization, despite the fact that the pumping of the system is homogeneous in both intensity and polarization.

\subsection{Spin Defects in the Dynamically Stable Regime}

\begin{figure}[tbh]
\includegraphics[width=1.0\linewidth]{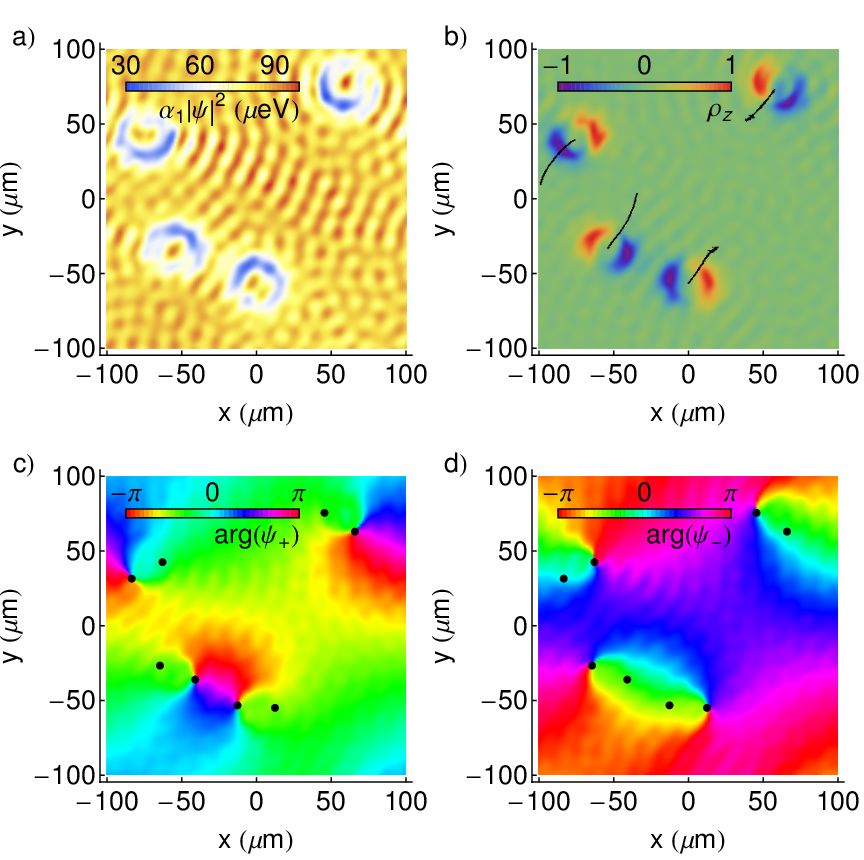}
\caption{(Color online) Spin defect formation without MI. a) Density pattern of a polariton condensate under homogeneous incoherent pumping. b) Distribution of the circular polarization degree, $\rho_z=\left(|\psi_+|^2-|\psi_-|^2\right)/\left(|\psi_+|^2+|\psi_-|^2\right)$. The black curves show the path traced out by the eyelets over $1000$~ps. c) and d) show the phase of the $\psi_+$ and $\psi_-$ polariton fields, respectively. Black spots indicate the positions of (half) vortices. Parameters were the same as in Fig.~\ref{fig:spinor} (with $P=84$~ps$^{-1}\mu$m$^{-2}$).}
\label{fig:spinpattern}
\end{figure}

When the system is excited with a larger pump power, the system tends to follow the stable branch (S) in Fig.~\ref{fig:spinor}. In this case one can expect a spatially HS due to the stability of the (S) branch, however, defects present in the initial state after transition to condensation (in simulations taken as a low intensity white noise as in sections ~\ref{sec:scalarnonlinear} and~\ref{sec:scalarvortices}) are trapped in the system and stabilize with the structure shown in Fig.~\ref{fig:spinpattern}.

The defects have a non-trivial density structure with localized maxima inside of an otherwise circular shaped drop in density. These ``eyelets'' are not fixed in their locations, but move randomly in the plane with a typical speed on the order of $0.05~\mu$m/ps. The eyelets also possess a characteristic spin polarization with a dipole type shape, as shown in Fig.~\ref{fig:spinpattern}(b). The structure of the eyelets can also be understood when looking at the phase distribution, which is shown in Figs.~\ref{fig:spinpattern}(c) and (d) for the $\sigma^+$ and $\sigma^-$ polarized components, respectively. Here it is clear that each eyelet is formed from a pair of half vortices~\cite{Rubo2007} -- one appearing in each spin component.

In addition to the slow drift, the eyelets preserve their shape while undergoing a faster periodic rotation with a period close to $2\pi/\Delta_{XY}$ (a movie of the motion is available in the Supplemental Material). Apparently this periodical motion of two bound vortices with opposite spins is induced by the polarization splitting in spinor condensates. It has been shown recently~\cite{Egorov2014b} that similar spinor effects can evoke a uniform motion of bound polariton solitons in coherently driven microresonators.


\subsection{Scaling Laws for the Defect Density} \label{Ch:ScalingLaw}

\begin{figure}[tb]
\includegraphics[width=1.0\linewidth]{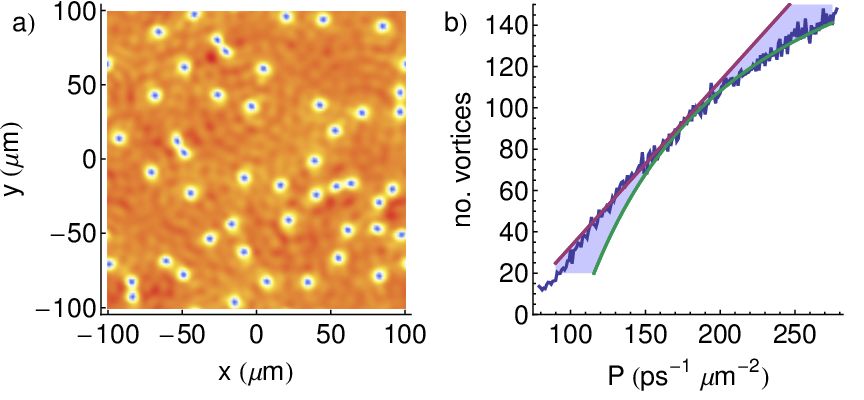}
\caption{(Color online) (a) Same as in Fig.~\ref{fig:spinpattern} under higher power excitation. (b) Variation of the number of vortices with pump power obtained numerically. The numerical results are averaged over ten different initial configurations at each value of pump intensity. The scaling with pump power is compared to the dependencies $N_d\sim P_{th} \left(1-P_{th}/{P}\right)$ (green curve) and $N_d\sim\left(P/P_{th}-1\right)$ (magenta curve), see Eqs.~(\ref{eq:KZ1}) and (\ref{eq:KZ2}).}
\label{fig:power}
\end{figure}

The number of vortices generated in the system outside of the MI region, during the mean-field evolution of an initially noisy, low-density state, is found to grow with the pump intensity, as shown in Fig.~\ref{fig:power}. The number of vortices is counted at a time after spatial coherence is established in the system, where the average polariton and reservoir density achieves a steady-state. Note that at very long-times there may be further recombination of vortex-antivortex pairs.

By taking advantage of the universality of dynamics of the system in the vicinity of the phase transition, we find approximate scaling laws governing the number of defects created during the transition,
see solid line in Fig.~\ref{fig:power}(b). The scaling laws in the case of exciton-polaritons have similar forms as the ones obtained from the argument of Kibble and Zurek.\cite{Kibble1976,Zurek1985}
However, the dynamics of defect formation is different due to the fact that the initial state of the system is strongly out of equilibrium.\cite{Matuszewski2014} While in the Kibble-Zurek mechanism the phase transition is assumed to begin in the initial state that is close to thermal equilibrium, here the white-noise initial state is dominated by fluctuations. In both cases, spontaneous symmetry breaking occurs differently in separate regions of space which cannot communicate on how the symmetry is broken due to the finite timescale of the process. On the borders between these separate regions defects can appear in the form of domain walls, vortices, or more sophisticated structures, depending on the dimensionality of the system and the form of the order parameter.\cite{Kibble1976,Zurek1985,Other_KZ}

In the case of dynamics described by (\ref{eq:GPplus})-(\ref{eq:Reservoir}), starting from the initial low density white-noise state, the growth of polariton density initially occurs without establishing any coherence (a movie showing the dynamics corresponding to the establishment of the state in Fig.~\ref{fig:power}a can be found in the supplemental material). This corresponds to the first stage of the process, where there is practically no nonlinearity and no $k$-dependence of the growth rate. The defects are created in the second stage, when fluctuations are suppressed due to nonlinear interactions. In this process, regions of ordered phase (or patches of a condensate) appear out of the initial disordered strongly fluctuating phase. Due to spatial and polarization symmetry breaking, defects are created on the borders between these condensed regions.

We assume that in the emerging ordered regions described above the relative phase between the two polarization components is approximately equal to $\pi$ and densities of $|\psi_+|^2$ and $|\psi_-|^2$ are similar. This assumption is in agreement with the numerical data presented in Figs.~\ref{fig:spinpattern}(c) and \ref{fig:spinpattern}(d), where the phase is equal to $\pi$ away from the eyelets, and is related to the fact that such configuration minimizes the energy in Eqs.~(\ref{eq:GPplus}-\ref{eq:GPminus}).
We substitute $\psi_{\pm}=\pm \frac{1}{\sqrt{2}}\psi$ to obtain
\begin{align}
i\hbar\frac{d\psi}{dt}&=\left(- \frac{\hbar^2\nabla^2}{2m}+g_1n_R+\frac{\alpha_1+\alpha_2}{2}|\psi|^2\right)\psi\notag\\
&\hspace{10mm}+\frac{i\hbar}{2}\left(r n_R-2\Gamma\right)\psi-\Delta_\mathrm{XY}\psi,\notag\\\label{eq:GP1}
\frac{dn_R}{dt}&=-\left(\Gamma_R+\frac{r}{2}|\psi|^2\right)n_R+P.
\end{align}
%
Next, we assume that the reservoir quickly adjusts to the change of $|\psi|^2$,
\begin{equation} \label{nR0}
n_R=\frac{P}{\Gamma_R + r|\psi|^2/2}.
\end{equation}
Let us now assume that there exists a patch of approximately constant polariton field $\psi_0$ (or a condensate seed) at $t=t_1$. We can describe small fluctuations around $\psi_0$ by
\begin{equation} \label{Bogoliubov}
\psi = \left(\psi_0 + u(t) e^{ikx} + v^*(t) e^{-ikx} \right) e^{-i\mu t + \lambda (t-t_1)}\,,
\end{equation}
where $\mu$ and $\lambda$ are the chemical potential and the growth rate of the patch, and $u$, $v$ represent the fluctuations. Expanding (\ref{nR0}) in Taylor series around $\psi_0$, we can rewrite (\ref{eq:GP1}) as
\begin{equation} \label{modelA}
i \hbar \dot {\psi} = \bigg[-\frac{\hbar^2 \nabla^2}{2 m}
+ i\frac{\hbar\gamma_0}{2} \left(1-\frac{|\psi|^2}{n_{\rm sat}}\right)
+g_{\rm eff} |\psi|^2 + \mu_{\rm a}  \bigg] \psi\,,
\end{equation}
with the accuracy of the order of $O(|\psi|^4)$. Here $g_{\rm eff}=(\alpha_1+\alpha_2)(1-g_1 r P/2\Gamma_A^2)/2$,
$\gamma_0 = (rP/\Gamma_A) (1+ r|\psi_0|^2/2\Gamma_A)-2\Gamma$,
and $n_{\rm sat}=2\gamma_0\Gamma_A^2/(r^2 P) $ where $\Gamma_A = \Gamma_R +  r |\psi_0|^2/2$ is the effective exciton decay rate.
The above equation has the same form as the one derived in Ref.~[\onlinecite{Matuszewski2014}], apart from the negligible dependence of $g_{\rm eff}$ on $P$.
The Bogoliubov-de Gennes modes are $u,v^* \sim e^{-i\omega_\pm t}$, and the mode frequencies
\begin{equation} \label{eq:GP1BdG}
\frac{\omega_\pm}{\gamma_0} = -i\frac{\alpha}{2} \pm i\sqrt{\left(\frac{\alpha}{2}\right)^2 +(\alpha \beta)^2 - \left(
\varepsilon_k+\alpha \beta \right)^2},
\end{equation}
where $\varepsilon_k=\hbar k^2 / 2 m \gamma_0$, the saturation parameter $\alpha=|\psi_0|^2 / n_{\rm sat}$, and $\beta=2g_{\rm eff}\Gamma_A^2/\hbar r^2P$.
This spectrum has the property that modes with high momenta are strongly damped (the imaginary part of the frequency is negative), in contrast to the initial linear dynamics where no $k$-dependence of the imaginary part of the spectrum was present. We can define a characteristic momentum cutoff $\kappa$ for the modes that are strongly damped, which scales
with the parameters as $\kappa\sim \gamma_0^{1/2}$. Any fluctuations with momenta higher than $\kappa$ will be suppressed, while fluctuations with lower momentum can form regions of ordered condensate phase.
To estimate the scaling of the number of defects, two different limiting cases can be considered. In the first case, we assume that the vortices are formed when the condensate density is already near to its equilibrium value $|\psi_0|^2=\left(P-P_{\rm th}\right)/\Gamma$. Here we obtain
\begin{equation}
N_d\sim \kappa^2 \sim \gamma_0 \sim P_{\rm th}\left(1 - \frac{P_{\rm th}}{P}\right).\label{eq:KZ1}
\end{equation}
Note that the above scaling does not have a power-law form, which is due to the fact that the transition is effectively nonlinear.

In the opposite limit,\cite{Matuszewski2014} we assume that the vortices are formed when the condensate density is still very small $|\psi_0|^2=0$. In this case we obtain
\begin{equation}
N_d\sim \kappa^2 \sim \gamma_0 \sim \frac{P}{P_{\rm th}}-1.\label{eq:KZ2}
\end{equation}
The two estimates [Eqs.~(\ref{eq:KZ1}) and (\ref{eq:KZ2})] are compared to the numerical results in Fig.~\ref{fig:power}(b). The numerically obtained scaling appears to be intermediate between the two extreme cases.

\section{Conclusions}

In this paper we presented a comprehensive theoretical study of non-equilibrium dynamics of polariton condensates in incoherently pumped semiconductor microcavities. We have anticipated two different destabilization mechanisms that govern nonlinear dynamics of this system. The first arises when the polariton condensate has a strong feedback effect on the reservoir of incoherent ``hot'' polaritons. The second one is associated with the parametric scattering in the presence of polarization splitting in a spinor condensate. Both mechanisms result in the formation of phase defects, i.e. vortices, triggered by the modulational instability of the homogeneous condensate.

In the scalar case, we have shown that the presence of the incoherent reservoir can affect substantially both the vortex stability and their mutual collective dynamics. In particular this can lead to the formation of rotating dislocations or delocalized spiraling waves.

In the spinor two-component condensate we have identified the presence of topological defects, which take the form of half-vortex pairs, giving an ``eyelet'' structure in intensity and dipole type structure in the spin polarization.

In the case when the phase defects are formed in the dynamically (modulationally) stable regime, as a result of the condensate formation from an initial spatially and phase-incoherent state, we find that the defect density scales with the pumping rate in analogy to the Kibble-Zurek type scaling for non-equilibrium phase transitions.

\section*{ACKNOWLEDGEMENTS}

T.C.H.L. acknowledges support from the Lee Kuan Yew Fellowship. O.A.E. acknowledges financial support by the Deutsche Forschungsgemeinschaft (DFG project EG344/2-1) and the Thuringian Ministry for Education, Science and Culture (TMESC project B514-11027). M.M.~acknowledges
support from the National Science Center grant DEC-2011/01/D/ST3/00482. E.A.O. acknowledges funding by the Australian Research Council (ARC). \\

\appendix

\section{Linear Stability Analysis}\label{Ch:AppendixStability}

The stability of the solutions can be checked using the standard approach of applying perturbations, in the form $\psi_\pm\mapsto e^{-i\mu t}\left(\psi_\pm+\delta\psi_\pm\right)$, $n_R\mapsto n_R+\delta n_R$. Substitution into Eqs.~(\ref{eq:GPplus})-(\ref{eq:Reservoir}) and collecting terms linear in the small amplitudes $\delta\psi_\pm$ and $\delta n_R$ we have:
\begin{align}
i\hbar\frac{d\delta\psi_+}{dt}&=c_1\delta\psi_++\left(\Delta_{XY}+\alpha_2\psi_-^*\psi_+\right)\delta\psi_-+\alpha_1\psi_+^2\delta\psi_+^*\notag\\
&\hspace{5mm}+\alpha_2\psi_-\psi_+\delta\psi^*_-+\left(g_1+i\hbar r\right)\psi_+\delta n_R,\label{eq:fluct1}\\
i\hbar\frac{d\delta\psi_-}{dt}&=c_2\delta\psi_-+\left(\Delta_{XY}+\alpha_2\psi_+^*\psi_-\right)\delta\psi_++\alpha_1\psi_-^2\delta\psi_-^*\notag\\
&\hspace{5mm}+\alpha_2\psi_+\psi_-\delta\psi^*_++g_2\psi_-\delta n_R,\label{eq:fluct2}
\end{align}
\begin{equation}
\frac{d\delta n_R}{dt}=-\left(\Gamma_R+r|\psi_+|^2\right)\delta n_R-2rn_R\Re e\left\{\psi^*_+\delta\psi_+\right\},\label{eq:fluct3}
\end{equation}
where we have defined:
\begin{align}
c_1&=g_1n_R+i\hbar \left(rn_R-\Gamma\right)+2\alpha_1|\psi_+|^2+\alpha_2|\psi_-|^2,\\
c_2&=g_2n_R-i \hbar\Gamma+2\alpha_1|\psi_-|^2+\alpha_2|\psi_+|^2.
\end{align}
%

We consider perturbations in the polariton and reservoir fields of the form $\delta\psi_\pm=u_\pm e^{i(kx-\omega t)}+v^*_\pm e^{-i(kx-i\omega^* t)}$ and $\delta n_R=w\left(e^{i(kx-\omega t)}+e^{-i(kx-\omega^* t)}\right)$, respectively. Here $u_\pm$ and $v_\pm$ are complex amplitudes, while $w$ is a real amplitude. $\omega$ is a complex eigenvalue to be determined. Stability of the original solutions occurs if $\Im m\left\{\omega\right\}<0$ such that the perturbation decays.

Substitution of $\delta\psi_\pm$ and $\delta n_R$ into Eqs.~(\ref{eq:fluct1})-(\ref{eq:fluct3}) and collection of terms oscillating as $e^{-i\omega t}$ and $e^{-i\omega^* t}$ yields a set of five coupled equations, which represent an eigenvalue problem for $\omega$. In matrix form:
\begin{widetext}
\begin{align}
&\left(
\begin{array}{ccccc}
c_1'-\hbar\omega&\alpha_1\psi_+^2&\Delta_{XY}+\alpha_2\psi_-^*\psi_+&\alpha^2\psi_-\psi_+&(g_1+i\hbar r)\psi_+\\
-\alpha_1\psi^{*2}_+&-c_1'^*-\hbar\omega&-\alpha^2\psi_-^*\psi_+^*&-\Delta_{XY}-\alpha_2\psi_-\psi_+^*&-(g_1-i\hbar r)\psi_+\\
\Delta_{XY}+\alpha_2\psi^*_+\psi_-&\alpha_2\psi_-\psi_+&c_2'-\hbar\omega&\alpha_1\psi_-^2&g_2\psi_-\\
-\alpha_2\psi_-^*\psi_+^*&-\Delta_{XY}+-\alpha_2\psi_+\psi_-^*&-\alpha_1\psi_-^{*2}&-c_2'^*-\hbar\omega&-g_2\psi_-^*\\
-i\hbar rn_R\psi_+^*&-i\hbar rn_R\psi_+&0&0&-i\hbar(\Gamma_R+r|\psi_+|^2)-\hbar\omega
\end{array}
\right)
\left(\begin{array}{c}u_+\\v_+\\u_-\\v_-\\w\end{array}\right)\notag\\&\hspace{15cm}=0,\label{eq:BogoliubovMatrix}
\end{align}
\end{widetext}
where $c_1'=c_1+\frac{\hbar^2 k^2}{2m}-\hbar\mu$ and $c_2'=c_2+\frac{\hbar^2 k^2}{2m}-\hbar\mu$.

\begin{figure}[t]
\includegraphics[width=1.0\linewidth]{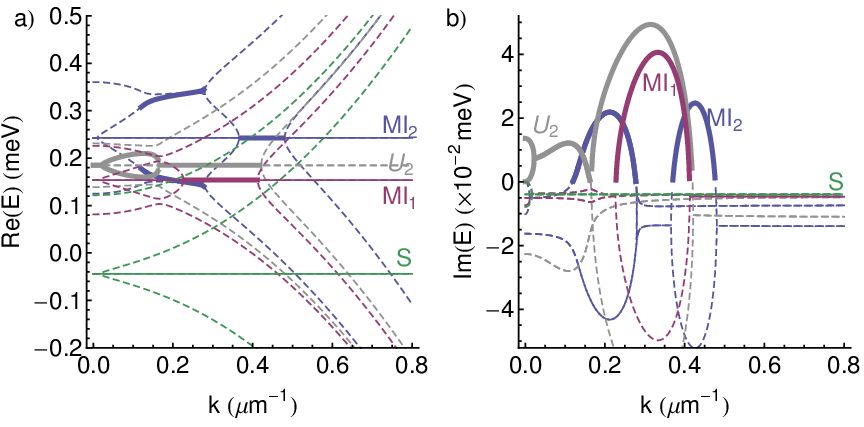}
\caption{(Color online) Real (a) and imaginary (b) parts of the spectrum of fluctuations $E=\hbar\omega$ about stationary solutions. Four stationary state solutions are considered, with energies $\hbar\mu$ marked by the thin horizontal lines in (a). When the imaginary parts of $E$ are positive, the corresponding stationary solution is unstable, with the instability region marked by thick solid curves. Parameters were the same as in Fig.~\ref{fig:spinor}.}
\label{fig:spinorspectrum}
\end{figure}

The imaginary and real parts of the perturbation spectra, $E=\hbar\omega$, are shown in Fig.~\ref{fig:spinorspectrum}(a,b). The energies of the stationary states $\hbar\mu$ are illustrated by thin horizontal lines in Fig.~\ref{fig:spinorspectrum}(a). Where the imaginary parts are positive, thick curves denote modulational instability at the given wavevectors $k$. The stationary state labelled $U_2$ is unstable to fluctuations with $k=0$.

\end{document}